\documentstyle[12pt,aaspp4,flushrt]{article}

\def\today{\ifcase\month\or January\or February\or March\or April\or May\or
  June\or July\or August\or September\or October\or November\or December\fi
  \space\number\day, \number\year}
\def\be{\begin{equation}}
\def\ee{\end{equation}}
\def\capt{\small \baselineskip 12pt }
\def\etal{{\it et al.\ }}
\def\rms{{\it rms\ }}

\def\eg{{\it e.g.}}

\def\apriori{{\it a priori\ }}
\def\ltsima{$\; \buildrel < \over \sim \;$}
\def\lsim{\lower.5ex\hbox{\ltsima}}
\def\gtsima{$\; \buildrel > \over \sim \;$}
\def\gsim{\lower.5ex\hbox{\gtsima}}
\def\kms{\ {\rm km\,s^{-1}}}
\def\hmpc{\,{\rm h^{-1}Mpc}}
\def\kmsmpc{\,{\rm km\,s^{-1}Mpc^{-1}}}

\def\log{{\rm log}}
\def\mag{{\rm mag\,}}

\def\omm{\Omega_m}
\def\oml{\Omega_\Lambda}

\def\omk{\Omega_k}

\def\V{\Phi}
\def\pmb#1{\setbox0=\hbox{#1}%
 \kern-.025em\copy0\kern-\wd0
 \kern.05em\copy0\kern-\wd0
 \kern-.025em\raise.0433em\box0}
\def\vv{\pmb{$v$}}

\begin{document}
 
\title{A LOCAL HUBBLE BUBBLE FROM SNe Ia?}
 
\author{Idit Zehavi \altaffilmark{1}, Adam G. Riess \altaffilmark{2}, \\
Robert P. Kirshner \altaffilmark{3} and Avishai Dekel \altaffilmark{1}}

\altaffiltext{1}{Racah Institute of Physics, The Hebrew University,
Jerusalem 91904, Israel} 
\altaffiltext{2}{Astronomy Department, University of California,
Berkeley, CA 94720}
\altaffiltext{3}{Harvard-Smithsonian Center for Astrophysics, 60 Garden St.,
Cambridge MA 02138} 

\begin{abstract}

We analyze the monopole in the peculiar velocities of 44 Type Ia
supernovae (SNe Ia) to test for a local void.  The sample extends from
$20$ to $300\hmpc$, with distances, deduced from light-curve shapes,
accurate to $\sim 6\%$. Assuming $\omm=1$ and $\oml=0$, the most
significant deviation we find from the Hubble law is an outwards flow
of $(6.6\pm2.2)\%$ inside a sphere of radius $70\hmpc$ as would be
produced by a void of $\sim 20\%$ underdensity surrounded by a dense
shell.  This shell roughly coincides with the local Great Walls.
Monte Carlo analyses, using Gaussian errors or bootstrap resampling,
show the probability for chance occurrence of this result out of a
pure Hubble flow to be $\sim 2\%$. The monopole could be contaminated
by higher moments of the velocity field, especially a quadrupole,
which are not properly probed by the current limited sky coverage. The
void would be less significant if $\omm$ is low and $\oml$ is high.
It would be more significant if one outlier is removed from the
sample, or if the size of the void is constrained \apriori. This
putative void is not in significant conflict with any of the standard
cosmological scenarios. It suggests that the Hubble constant as
determined within $70\hmpc$ could be overestimated by $\sim 6\%$ and
the local value of $\Omega$ may be underestimated by $\sim 20\%$.
While the present evidence for a local void is marginal in this data
set, the analysis shows that the accumulation of SN Ia distances will
soon provide useful constraints on elusive and important aspects of
regional cosmic dynamics.

\end{abstract}
Subject headings: Cosmology: observations --- cosmology: theory --- 
                  galaxies: distances and redshifts --- large-scale 
                  structure of universe --- supernovae: general
\vfill
\eject

%==============================================================
\section{INTRODUCTION}
%1
\label{sec:intro}

%Preliminary evidence for a local void and its scale from redshift surveys:
Large-scale redshift surveys of galaxies show underdense regions of
typical extent $\sim 50\hmpc$. These ``voids" appear to be bordered by
dense ``walls" (Kirshner \etal 1981; Huchra \etal 1983 [CfA];
Broadhurst \etal 1990; Shectman \etal 1996 [LCRS]). In particular,
maps of our cosmological neighborhood display the Great Wall of Coma
and the Southern Wall, that appear to connect into a shell-like
structure of radius $70-80\hmpc$ about the Local Group (Geller \&
Huchra 1989 [CfA2]; da Costa \etal 1994 [SSRS2]). The volume
encompassed by this structure appears to be of lower density.
Despite these impressive maps, it is difficult to quantify the
large-scale radial density profile of this region. First, the true
galaxy density is hard to distinguish from the sample selection
function when the structure of interest approaches the sample size.
Second, we do not know how well galaxies trace mass. And third,
portions of the galaxy distribution are obscured from our viewpoint
within the Milky Way.
The imprint of wall and structure on the power spectrum may possibly
be associated with excess power observed at a wavelength of $\sim
100-150\hmpc$ (\eg, Broadhurst \etal 1990; Landy \etal 1996; Einasto
\etal 1997). This scale might be naturally attributed to the scale of
the cosmological horizon after the universe became matter dominated
and before the plasma recombined, though this peak is only expected to
be significant for relatively high values of the baryon density (Hu \&
White 1997).

%General motivation for a local void
A local void has been proposed as one way to reconcile the age of the
Universe based on the Hubble expansion with the ages of globular
clusters within the framework of the Einstein-de Sitter cosmology
(\eg, Turner, Cen \& Ostriker 1992; Bartlett \etal 1995) Measurements
of the Hubble constant within the void would overestimate the
universal value by $\delta \rho/\rho \approx -3\, \delta H/H$.
Indeed, the values obtained for the Hubble constant from the
longest-range distance indicators, the SNe Ia (Jacoby et al 1992;
Sandage \& Tammann 1993; Tammann \& Sandage 1995; Hamuy et al 1995,
1996b; Riess, Press, \& Kirshner 1995a, 1996; Branch, Nugent, \&
Fisher 1996) and the gravitational lenses (Falco \etal 1997; Keeton \&
Kochanek 1997) are typically smaller than values obtained more locally
using Tully-Fisher (TF) distance indicators (Kennicutt, Freedman, \&
Mould 1995; Mould \etal 1995; Freedman \etal 1994; Freedman 1997).
A local void would also imply that local estimates of $\Omega$
underestimate the global value of $\Omega$.
Finally, a local outflow would reduce the distances derived from TF
peculiar velocities for features such as the Great Attractor, bringing
them into better agreement with the positions derived from redshift
surveys (Sigad \etal 1998).

%No detection so far
It is important to separate impressions and theoretical wishes from
firmly established observational facts.  Attempts to establish
monopole deviations from pure Hubble flow have not yet produced
conclusive results. For example, Shi (1997) claimed finding a monopole
gradient in subsamples of the Mark III catalog of Tully-Fisher
peculiar velocities (Willick \etal 1997a).  But he also found a
marginal monopole gradient of the opposite sign in the peculiar
motions of rich clusters based on their brightest members as standard
candles (Lauer \& Postman 1992; 1994). Finally, he found no
significant deviation in an early subsample of 20 SNe (Riess \etal
1996). None of these data sets was ideal for testing for a void of
radius $\sim 70\hmpc$: the Mark III data include only a small number
of galaxies beyond $70\hmpc$, the earlier SNe sample is too sparse,
and the rich clusters, while of comparable abundance to the current
SNe sample within $70\hmpc$, have a much larger error per cluster.
Kim \etal (1997), based on comparing SNe distances at high redshifts
($z\sim 0.4$) and low redshifts ($z < 0.1$), put a $2\sigma$ upper
limit of $10\%$ for an outflow within the local $300\hmpc$, but their
sample has only a few SNe within $70\hmpc$, which is the suggested
domain of the putative void.

We now have a sample of 44 SNe Ia from the Calan/Tololo Survey (Hamuy
et al 1993, 1996a) and from the CfA supernova program (Riess 1996,
Riess \etal 1998), which reaches the threshold for an interesting
assessment of the dynamical signature of a void contained within the
great walls. The nearest SNe in this sample are at distances of $\sim
20\hmpc$, the furthest are beyond $300\hmpc$, the median is at $\simeq
95\hmpc$, and 17 SNe lie inside $70\hmpc$.  This distribution in
distance is well suited for detecting a monopole perturbation within
$50-100\hmpc$ while determining the universal Hubble flow outside this
sphere. These are the scales which have been suggested and where the
standard cosmological theories predict perturbations on the order of
several percent (\eg, Wang, Spergel \& Turner 1998).  Because SNe
light curves provide distances with a precision of $5-8\%$, we believe
it is worth carrying out a preliminary analysis with only 44
tracers. This sample has already been used to put constraints on the
global Hubble flow and the dipole motion (Hamuy et al 1995, 1996b;
Riess, Press, \& Kirshner 1995a,b, 1996), and the nearby sample has
been used to assess how well the velocities of SNe match the gravity
field inferred from galaxy redshift surveys (Riess \etal 1997).

In \S~\ref{sec:data} we describe the data.  In \S~\ref{sec:monopole}
we measure the peculiar monopole and evaluate its significance in an
Einstein-de Sitter Universe.  In \S~\ref{sec:cautions} we explore the
effect of different cosmologies, error estimates and higher
velocity-field moments on our results.  In \S~\ref{sec:discussion} we
discuss the implications of our results.

%=============================================================
\section{DATA}
%2
\label{sec:data}

% SNe distances
Our data consists of 44 SNe Ia light curves with distances ($d$)
inferred by the method of Multicolor Light Curve Shapes (MLCS) (Riess,
Press \& Kirshner 1996, RPK) and redshifts ($z$) (Hamuy \etal 1996a;
Riess 1996; Riess \etal 1998). The MLCS method can measure distances
with $\sim 6\%$ precision by accounting for variations in SNe Ia
luminosity and correcting for line-of-sight extinction.

% Curvature effect
The inferred distance is a ``luminosity" distance. To account for
cosmological curvature effects we define the peculiar velocity as
$u=d_l(z) - d$, where $d_l(z)$ is the luminosity distance (in units of
$\kms$) corresponding to redshift $z$ for the assumed values of the
cosmological parameters $\omm$ and $\oml$.  It is given by
\begin{equation} 
d_l(z) = {c (1+z)\over \vert\omk\vert^{1/2} }\
S_k \left[\,\vert\omk\vert^{1/2} \int_0^z dz' \,
[(1+z')^2 (1+\omm z') -z'(2+z')\oml ]^{-1/2}
\right] \ ,
\label{eq:d_l}
\end{equation}
where $k$ is the curvature parameter, $\omk=1-\omm-\oml$,
$S_0(x)\equiv x$, $S_{+1}\equiv \sin$ and $S_{-1}\equiv \sinh$.  Every
$z$ hereafter actually refers to $d_l(z)$.

% Errors uncertainties
``Formal" estimates of the distance errors for the individual SNe,
including the intrinsic scatter in the distance indicator and the
measurement errors, are provided by the covariance matrix of the MLCS
fit (RPK), $\Delta_{MLCS}$. These internal errors are typically $5-8\%$
of the distance.  Additional systematic errors in the distances are
uncertain, so we adopt a conservative approach by adding in quadrature
$\Delta_{add}=0.05$ \mag in our ``standard" study case (in
\S~\ref{sec:monopole}), and later discuss the implications of varying
$\Delta_{add}$ in the range $0-0.1$ \mag.

%===============================================================
\section{PECULIAR MONOPOLE}
%3
\label{sec:monopole}

%--------------------------------------
\subsection{Fitting a Monopole Profile}

\begin{figure}[t]
%1
{\includegraphics{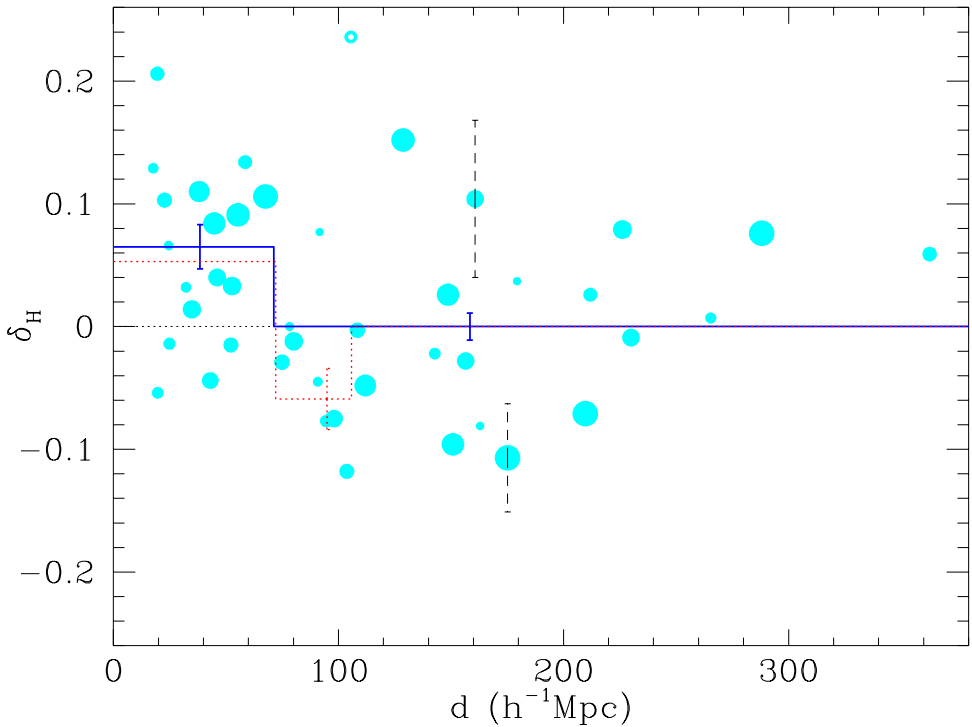}}
\vspace{9.5 cm} 
\caption{\capt
The monopole profile: $\delta_H$ versus inferred distance $d$. The
individual SNe are marked as circles of radii inversely proportional
to the errors (the actual errors are shown for 2 representative SNe to
calibrate the size of the symbols).  Overlayed are the best-fit 2-zone
(solid) and 3-zone (dashed) monopole models, together with the
corresponding errors of the fits. The global $H$ is determined from
the outer region. The outlier 92bl is marked by white dot.  }
\label{fig:dH}
\end{figure}

Figure~\ref{fig:dH} shows the radial peculiar velocities of the 44
SNe, $u$, translated to the corresponding relative deviations from a
pure Hubble flow, $\delta_H\equiv \delta H/H = u/d$, at the inferred
distances $d$, assuming (in this section) $\Omega=1$ and $\Lambda=0$.
The plot suggests positive $\delta_H$, an outflow, inside $70\hmpc$,
and possibly a marginal indication for an inflow in the surrounding
shell $70-110\hmpc$, before the flow converges to a global Hubble
expansion at larger distances.

% Outlier
The 26th SN from the Local Group at $d\simeq 107\hmpc$, named SN
1992bl, is the most discrepant point in this plot.  Its inferred
peculiar velocity of $\sim 2400 \kms$ is a $\sim 4\sigma$ deviation
from the global Hubble flow. As described below, the exclusion of this
outlier from the analysis makes a difference in the significance of
the results.  Though we have no reason to think that SN 1992bl is
incorrectly analyzed, and we have not found anything obvious that
makes it an a priori suspect, we want to evaluate the impact of this
single point on the significance of the void, and we therefore report
the results both with and without this outlier.

While visual inspection of Fig.~\ref{fig:dH} suggests the possibility
of a local void, our approach is to use a statistical analysis to
provide a quantitative estimate of the significance of the void.
%
% H by chi2 minimization
The Hubble constant $H$ within a given volume is determined by
minimizing the error-weighted sum of residuals over the SNe,
\begin{equation}
\chi^2 = \sum_i [\log(z_i/d_i) - \log H]^2 /\sigma_i^2 ,
\label{eq:chi2}
\end{equation}
where $\sigma_i$ is the error in $\log(z_i/d_i)$.  The best-fit Hubble
constant and the corresponding variance are then
\begin{equation}
\log H = \sigma^2 \sum \sigma_i^{-2} \log (z_i/d_i) ,
\quad \sigma^2 = (\sum \sigma_i^{-2})^{-1} .
\label{eq:H}
\end{equation}
 
The scatter about a pure Hubble flow results from both distance
uncertainties and peculiar motions.  We assume that the distance error
is distributed Normally in $\log d$, with a standard deviation
$\Delta$, where $\Delta$ is the sum in quadrature of the formal
distance error of the given SN and the additional scatter
$\Delta_{add}=0.05 \mag$ adopted as our ``standard" case
($\Delta_i^2=\Delta_{MLCS}^2+\Delta_{add}^2$).  For the peculiar
velocities we assume an \rms value, $\sigma_f$ (in frame of the Cosmic
Microwave Background, CMB), independent of distance.  For SNe in the
sample beyond $\sim 3000\kms$, the distance error dominates the
peculiar velocity uncertainty.  Since only six SNe are closer, we
approximate the {\it total\,} scatter of $\log (z/d)$ as Normal, with
a variance $\sigma_i^2 = \Delta_i^2 + (\sigma_f/z_i)^2$.
 
Partly motivated by the low dispersion velocity detected for galaxies
in redshift and peculiar velocity surveys (Davis \& Peebles 1983;
Willick \etal 1997b; Davis, Miller \& White 1997), we adopt
$\sigma_f=200\kms$, and explore other values later.  This value for
$\sigma_f$ provides a $\chi^2$ per degree of freedom of $\chi^2_{dof}
= 0.99$ within the inner $100\hmpc$, where the data and the formal
error estimates are most reliable and where $\sigma_f$ makes a
difference.  The contribution of SNe in the outer regions makes the
overall value somewhat larger, $\chi^2_{dof} = 1.9$, which suggests
that we may underestimate the errors for a few SNe at very large
distances.

%------------------------------
\subsection{1-2-3-Zone Models}

%Fitting 1-2-3 zone models
We try to fit the data with three alternative simple models and
compare the corresponding $\chi^2$.  (a) a 1-zone no-void model of
uniform expansion with Hubble constant $H$, (b) a 2-zone void model
with a local Hubble constant $H_1$ within a sphere of radius $R_1$,
and a global Hubble constant $H$ outside of this sphere, and (c) a
3-zone void model with an additional shell between the radii $R_1$ and
$R_2$ within which the Hubble constant is $H_2$.  These models have 1,
3, and 5 free parameters respectively.
The model parameters are determined by minimizing the $\chi^2$
(eq.~\ref{eq:chi2}).  To maintain at least six SNe in each zone, $R_1$
and $R_2$ are limited by the current data to the range $30-200\hmpc$,

%Best-fit models
The best fit models, and the corresponding errors, are plotted in
Fig.~\ref{fig:dH}.  For the 2-zone model, the best-fit outflow zone
extends out to $R_1= 70\hmpc$ and the peculiar monopole in it is
$\delta_H=0.065\pm0.021$, where the error here includes in quadrature
the uncertainties in $H_1$ and $H$.  If 92bl is excluded, the outflow
is somewhat stronger, $\delta_H = 0.075 \pm 0.022$.  The location of
$R_1$ is extremely robust.
When allowing for a third, intermediate zone, the boundary of the
inner zone remains robust at $R_1=70\hmpc$, with
$\delta_H=0.053\pm0.022$, and it is surrounded by an inflowing shell
of radii $70-105\hmpc$, within which the inflow amounts to
$\delta_H=-0.059\pm0.027$.  (If 92bl is excluded, the extent of the
outflow zone would be the same, with $\delta_H=0.060\pm0.023$, and the
inflow shell would be wider, $70-120\hmpc$, and somewhat shallower,
$\delta_H=-0.048\pm0.025$.)

%2-zone and 3-zone models provide BETTER chi2 fits (1-3-5 params).
The values of $\chi^2$ that correspond to these model fits suggest
that the monopole deviation from a pure Hubble flow is {\it
significant}.  We obtain $\chi^2 = 80.9, 71.5, 67.1$ for the three
models respectively. This means a reduction of $4.7$ for each
additional parameter going from the 1-zone to the 2-zone model, and an
additional reduction of $2.2$ per parameter when moving from the
2-zone to the 3-zone model.  A reduction in $\chi^2$ of more than
unity per parameter is commonly regarded as a significant improvement.
Yet, because we had a chance to inspect the data before we developed
our model, one may argue that only a larger reduction should be
considered significant.  In addition, the fact that the $\chi^2_{dof}$
at very large distances is somewhat larger than unity complicates the
interpretation of the $\chi^2$ statistics. In any case, the detected
decrease in $\chi^2$ indicates that the 2-zone model is a significant
improvement over the pure Hubble flow model and should be taken
seriously, while the 3-zone model provides only marginal additional
improvement and stretches beyond the statistical reach of the current
data set.

%Without the outlier
If 92bl were excluded, the corresponding $\chi^2$ would be $67.5,
55.2, 51.5$ respectively. The improvement associated with the 2-zone
model would become even more pronounced, with the $\chi^2$ being
reduced by $6.1$ for each additional parameter. The improvement for
the 3-zone model is still marginal, of $1.9$ per each additional
parameter.
The exclusion of 92bl by itself would lead to significantly lower
$\chi^2$ values, by $13.4$ in the 1-zone model and by $16.3$ for the
2-zone model. A jackknife test of excluding one SN at a time and
redoing the fits confirms that 92bl is indeed the most deviant data
point.  However, it does not represent an abnormal deviation from a
Normal distribution.

%--------------------------------
\subsection{The Significance of a Void Within $70\hmpc$}

% R fixed a priori
Determining the significance of the peculiar monopole implied by the
model fits above is not trivial, and we therefore address it in
several different ways.  One way to assign a more concrete
significance is by fixing the radius of interest \apriori at
$R_1=70\hmpc$, based on the external indications for a local void
encompassed by the great walls (\S~\ref{sec:intro}), or on the
robustness of $R_1$ in the fits to the SNe data themselves.
The statistical question that we pose in this case is: at what
confidence level can one reject the null hypothesis that chance alone
could produce a fluctuation as large as we see inside $R_1=70\hmpc$,
while the underlying field is a pure Hubble flow. A crude answer can
be provided by the monopole deviation within $R_1$ in units of the
corresponding error, the ``voidness" $\V \equiv \delta_H / \sigma$.
We obtain $\V=3.0$ and $3.5$ with and without 92bl respectively.
Thus, for $R_1=70\hmpc$ fixed by assumption, the $6-7\%$ outflow is
significant at the $3\sigma$ level.

% comment about reduced sigf_in: 
The assumed value for the velocity dispersion in the inner void region
could be decreased once the data have been fit for a local monopole.
The global value of $\sigma_f$ was intended to include the
contribution from the monopole deviation, which has now been
explicitly removed from the velocity error budget within the local
void.  With a reduced value of $\sigma_f=135\kms$ that yields
$\chi^2_{dof}=1$ in the inner zone (based on scatter about the best
fit outflow model), the outflow within $70\hmpc$ becomes even more
significant: $\V=3.4$ and $3.9$ with and without 92bl respectively.
This may be a better estimate of the errors about the best-fit model,
but to keep our analysis conservative, we do not employ these reduced
errors in the estimates of significance that follow.

%------------------------------
\subsection{Significance by Gaussian Monte Carlo Analysis}

\begin{figure}[t]
%2
{\includegraphics{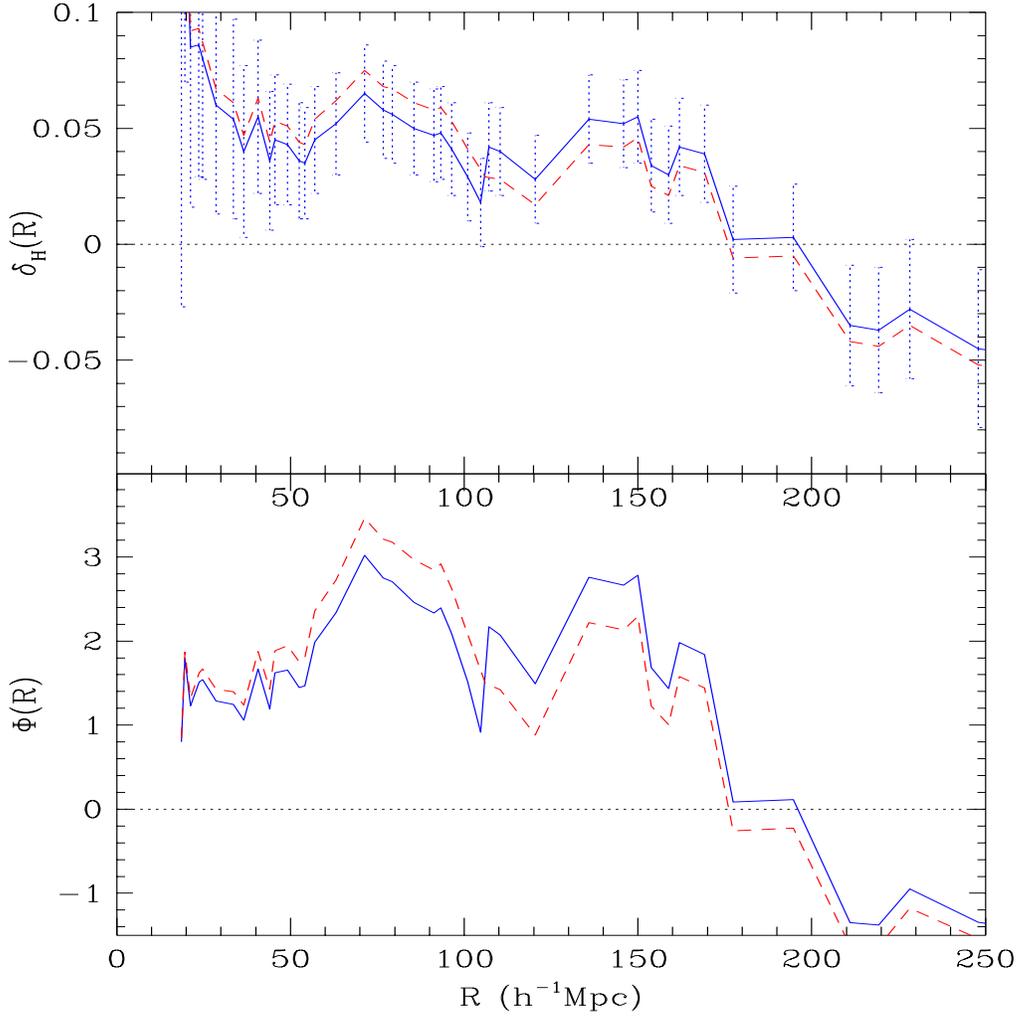}}
\vspace{13.2 cm} 
\caption{\capt
Top panel: $\delta_H(R)$ versus radius $R$, for all 44 SNe (solid)
with the corresponding error-bars, and when excluding the outlier
(dashed). $H$ is determined externally to $R$. Bottom panel: $\V(R)$
versus radius $R$, for all the 44 SNe (solid) and without the outlier
(dashed).}
\label{fig:V}
\end{figure}

% R determined from SNe data
An alternative, more general approach would be to let the size of the
void vary in a given range, then determine the extent of the most
significant void from the SNe data themselves, and estimate the
probability to observe a void of such depth (of any size in that
range).  The void size could be determined by minimizing the $\chi^2$
of a model that includes $R_1$ as a free parameter.
Another way to determine the extent and depth of the fluctuation in
the monopole (in the 2-zone model) is to search for the maximum of
$\V(R) \equiv \delta_H(R) / \sigma(R)$ over $R$ in a search range
$[r_{min},r_{max}]$; $\V$ is now defined as the maximum value of
$\V(R)$, and $R_1$ is the corresponding $R$.
Figure~\ref{fig:V} shows $\delta_H(R)$ with its error-bars (top panel)
and the resulting $\V(R)$ (bottom panel), both with and without 92bl.
The most pronounced monopole perturbation found in this way
corresponds to an outflow region of radius $R_1=70\hmpc$.
 
% A note about the 150 void:
There is apparently another feature at $R \sim 150\hmpc$, whose
relative significance decreases when 92bl is excluded.  This could be
an artifact from noisy determination of the universal Hubble constant,
since only 13 SNe lie outside this volume. When determining $H$
instead from all the data points, the value of $\V(R)$ at $R \sim
150\hmpc$ drops to about $60\%$ of its value at $70\hmpc$.

% Probability that the void is a fluke.  Gaussian MC analysis
When allowing the void size to be free, the value of $\V$ cannot be
interpreted simply as the significance of the void, because the radius
$R$ has been chosen to maximize $\V$.  Instead, we employ a Monte
Carlo analysis.  We generate random synthetic data sets with a single
zone of uniform expansion and with the same noise as in our real data.
We then compute the statistic $\V$ for each realization while allowing
$R$ to vary in a given search range, construct the probability
distribution function of $\V$, and see how far in the tail of this
distribution the observed maximum $\V$ lies. The corresponding
percentile, $p$, is the probability that the void is a chance
occurrence.  (This probability refers only to outflows of voidness
$>\V$, not including the opposite tail of the distribution of inflows
$<-\V$.)
The maximum search range allowed by the data is roughly $30-200\hmpc$,
which imposes only a weak initial constraint on $R$.  We may also
choose to try a sequence of more limited search ranges, all the way to
a tight range about $R=70\hmpc$, which would bring us back to the
alternative approach of focusing attention on a pre-determined scale.

% The MC simulations
% Gaussian MC
10,000 Monte Carlo realizations of the data are created by perturbing
a pure Hubble flow with a Gaussian distribution of errors.  The
synthetic SNe are initially placed at the redshift-space positions of
the 44 observed SNe. The distance moduli are then perturbed by a
Gaussian random variable of standard deviation equal to the assumed
distance error of that SN, and $z$ is perturbed further by a Gaussian
of width $\sigma_f$ to account for peculiar velocities.  The Monte
Carlo realizations aim at reproducing the same biases that arise in
the analysis of the real data due to coupling between the distance
errors and the deviations from Hubble flow.

For the most general search, allowing $R$ to be anywhere in the range
$30-200\hmpc$, we find $p=0.022$, that is, there is a $2.2\%$
likelihood that the void is a chance occurrence.  (When excluding
92bl, this probability reduces to $p=0.5\%$.) When the search is
limited to the intermediate range $50-100\hmpc$, the probability
becomes $0.9\%$ (and $0.2\%$ without 92bl).
 
The Gaussian synthetic catalogs are straightforward analogs of the
real data, but their validity relies on the accuracy of our error
estimates and their distribution.  The fact that the observed $\chi^2$
per degree of freedom is of order unity for the pure Hubble flow model
and our assumed errors, indicates that our assumed errors are not
unrealistic.
The probability distribution of the observed $\log (z/d)$ is nearly
indistinguishable from a Normal distribution, except, perhaps, for the
marginal outlier 92bl.  This suggests that our synthetic catalogs
properly represent the noisy data and that the corresponding
significance estimate is a useful guide. In \S~\ref{sec:cautions}, we
briefly explore the effect of using smaller or larger errors.

%--------------------------------------
\subsection{Error Estimation by Bootstrap Resampling}
% Bootstrap

Another way to evaluate significance that does not rely so heavily on
an assumed knowledge of the errors is to estimate the standard
deviation in the statistic $\V$ by bootstrap resampling of the noisy
data themselves, and using that estimate to evaluate the deviation of
the observed $\V$ from the null hypothesis of $\V=0$. (We do not use
the bootstrap realizations to derive the whole probability
distribution, the way we did with the Gaussian realizations, because
it could be biased.)
We create 10,000 synthetic catalogs by randomly selecting 44 data
points from the real data, allowing for repetitions.  We then
calculate the $\V$ statistic for each realization, and adopt the
standard deviation over the realizations as $\sigma_\V$.

We note that this bootstrap analysis is not completely independent of
the assumed errors, as they enter explicitly in the definition of
$\V(R)$ and in the $\chi^2$ fitting. However, because the realizations
are based on the real noise, this provides a partly independent way to
test the significance of the void.

The error estimated this way is $\sigma_\V = 1.2$.  It is slightly
larger than unity because $R$ is chosen in each realization to
maximize $\V$.  This value of $\sigma_\V$ implies that the confidence
level by which the observed $\V=3.0$ is different from $\V=0$ is $\sim
2.5\sigma$, or that the probability of obtaining such a {\it void\,}
by chance is roughly $0.6\%$.  (If 92bl is excluded, the result would
be $\V=3.5 \pm 1.1$, which can be similarly interpreted as a
$3.1\sigma$ effect, of $0.1\%$ probability.)

The bootstrap realizations show good evidence for significant voids at
values of $R$ near $70\hmpc$, as well as near $150\hmpc$ (see
Fig~\ref{fig:V}), because each of these regions contains many data
points that indicate a real effect.  The void significance based on
the bootstrap analysis is therefore similar to the significance found
from the Gaussian synthetic catalogs when the search range of $R$ is
narrow about the characteristic scale of the observed data, and higher
than the significance found when the search in $R$ is wide.

% Some comment about the error-bar of R 
The bootstrap analysis can also help us evaluate the robustness of the
void size. Since the $R$ of maximum $\V$ in the bootstrap realizations
sometimes occurs near $150\hmpc$ rather than near $70\hmpc$, the error
in the radius of the $70\hmpc$ void is hard to estimate when the
search range includes the two characteristic scales.  When the
$150\hmpc$ vicinity is excluded from the search range, the bootstrap
distribution yields $R=70 \pm 7 \hmpc$.  The uncertainty is set by the
coarseness of our data sampling and is roughly equal to the distance
between the two supernovae closest to $70\hmpc$.

%==================================================================
\section{CAUTIONS}
%4
\label{sec:cautions}

%--------------------
\subsection{Other Cases}

% Table 1
Table 1 summarizes the results of applying the tests described in 
\S~\ref{sec:monopole} to several other cases where we vary 
the cosmological parameters, the search range of void extent, the
inclusion of the outlier 92bl, and the assumed distance errors.
The cosmological density parameters $\omm$ and $\oml$ enter via the
``luminosity'' distance (eq.~[\ref{eq:d_l}]), which is approximately a
function of the deceleration parameter, $q_0=\omm/2-\oml$. The outlier
92bl is alternatively included or excluded. We test two options for
the search range of $R$: wide (30,200), and narrow (60,80).  The input
errors vary between the formal errors of the MLCS method
($\Delta_{add}=0$) and a generous overestimate of the errors, with
$\Delta_{add}=0.1$\mag.  The corresponding values of $\sigma_f=250$
and $120\kms$ respectively were obtained such that $\chi^2_{dof} \sim
1$ within the inner region (compared to a pure Hubble flow).
 
Listed are the resulting values of $\delta_H$ and $\V$ corresponding
to the best-fit inner zone (independent of the search range), the
error in $\V$ based on the bootstrap resampling, $\sigma_\V$, and the
(1-sided) probability $p$ that the void is a chance occurrence based
on the Gaussian Monte Carlo realizations.

\begin{table}[tbp]
%\vspace{-3 cm}
\caption{\capt
2-zone model results for the various cases} 
\vspace{0.5 cm}
\begin{tabular}{cccccccccc}
\tableline \tableline
 $\Delta_{add}$ & $\Omega_m$ & $\Omega_{\Lambda}$ & $92bl$ & 
$r_{min}-r_{max}$ & & \quad $\delta_H$ $\pm\sigma(\%)$ & $\V$ & 
$\pm \sigma_\V$ & Gaussian $p(\%)$ \\
\tableline

\vspace{-0.2cm}
0.05 &   1     &    0     &  in & 30-200 & & 6.5 $\pm$2.2 & 3.0 & 1.2 & 2.2 \\
%     &         &         &     & 50-100 & &              &     & 1.2 & 0.9 \\
     &         &         &     & 60-80  & &              &     & 1.2 & 0.5 \\
\vspace{-0.2cm} 
     &         &         & out & 30-200 & & 7.5 $\pm$2.2 & 3.5 & 1.1 & 0.5 \\
%     &         &         &     & 50-100 & &             &     & 1.2 & 0.2 \\
     &         &         &     & 60-80  & &             &     & 1.2 & 0.1 \\
\tableline
\vspace{-0.2cm}  
0.05 &    0.4  &   0     &  in & 30-200 & & 5.9 $\pm$2.1 & 2.8 & 1.2 & 4.3 \\
     &         &         &     & 60-80  & &             &     & 1.2 & 1.2 \\
\vspace{-0.2cm}
     &         &         & out & 30-200 & & 6.9 $\pm$2.2 & 3.2 & 1.2 & 1.2 \\
     &         &         &     & 60-80  & &             &     & 1.2 & 0.3 \\
\tableline
\vspace{-0.2cm}     
0.05 &    0.5  &   0.5   &  in & 30-200 & & 5.1 $\pm$2.1 & 2.4 & 1.2 & 9.8 \\
     &         &         &     & 60-80  & &             &     & 1.2 & 3.0 \\ 
\vspace{-0.2cm}
     &         &         & out & 30-200 & & 6.1 $\pm$2.2 & 2.8 & 1.1 & 3.5 \\
     &         &         &     & 60-80  & &             &     & 1.2 & 0.9 \\
\tableline
0.05 &    2    &   0     &  in & 30-200 & & 7.5 $\pm$2.2 & 3.5 & 1.2 & 0.6 \\
     &         &         & out & 30-200 & & 8.5 $\pm$2.2 & 3.9 & 1.1 & 0.1 \\
\tableline
 0   &    1    &   0     &  in & 30-200 & & 6.6 $\pm$2.2 & 3.0 & 1.4 & 1.8 \\
 0.1 &    1    &   0     &  in & 30-200 & & 6.3 $\pm$2.3 & 2.7 & 1.0 & 6.5 \\
\tableline

\end{tabular}
\end{table}

% Results
The derived size of the void is insensitive to the choice of
parameters in Table 1. The void in all cases encompasses the first
$17$ SNe and is of radius $R \simeq 70 \hmpc$.  The global Hubble
constant $H$, determined externally to the void, is robust at $\sim
65\pm 1 \kmsmpc$. (The error refers only to the $\chi^2$ fit, and the
sensitivity to the parameters that vary in Table 1 is even
smaller. This value of the Hubble constant is based on the original
calibration used by RPK, before Hipparcos.)
% Limited R range:
Constraining the search range for $R$ has the effect of increasing the
void significance based on the Gaussian Monte-Carlo realizations,
\eg, from $p=2.2\%$ to $0.5\%$ for our standard case. 
The standard deviation $\sigma_\V$ as estimated by the bootstrap resampling 
is less sensitive to the search range.
% Outlier:
The exclusion of the outlier 92bl increases the significance of the
void in all cases.

% Lower \Omega
Although most of the SNe are within $z<0.1$, the cosmological effects
are important because we are dealing with a small fluctuation of a few
percent.  Making no cosmological correction is equivalent to assuming
$\omm=2$ and $\oml=0$.
The depth and significance of the void both decrease gradually for
smaller $q_0$. The peculiar monopole $\delta_H$ goes down from $7.5\%$
(with no cosmological correction) to $5\%$ (for
$\Omega_{m}=\Omega_{\Lambda}=0.5$).  In the case
$\Omega_{m}=\Omega_{\Lambda}=0.5$, as long as we keep the outlier in
and perform a wide search for $R$, the void becomes only marginally
significant, with $p$ as high as $\sim 9.8\%$. However, even in this
case, the void becomes more significant, with $p=3-3.5\%$, if one
either excludes the outlier or limits the search range.

% \Delta_{add}=0;0.1
If one tries the formal errors only, the significance based on the
Gaussian synthetic catalogs becomes slightly larger than in the
standard case studied above; $p=1.8\%$ for $\Delta_{add}=0$ compared
to $2.2\%$ for $\Delta_{add}=0.05$.  Increasing $\Delta_{add}$ to
$0.1$ makes the significance weaker, $p=6.4\%$.  The significance is
increased back to the $p\sim 2\%$ level by either excluding the
outlier or limiting the search range.
The response of evaluation based on bootstrap to the assumed errors is
more complicated because these errors enter $\V$ and $\chi^2$. The
value of $\V/\sigma_\V$ can be crudely translated to a 1-sided
probability for a chance occurrence by assuming a Gaussian
distribution, yielding $p=1.8\%$ with $\Delta_{add}=0$ and $p=0.3\%$
for $\Delta_{add}=0.1$.  The fact that the bootstrap evaluation for
$\Delta_{add}=0.1$ indicates a considerably higher significance than
obtained from the Gaussian synthetic catalogs suggests that the errors
assumed in the Gaussian model are too large.

%--------------------------
\subsection{Higher Multipoles}

\begin{figure}[t]
%3
\vspace{0.5cm}
{\includegraphics{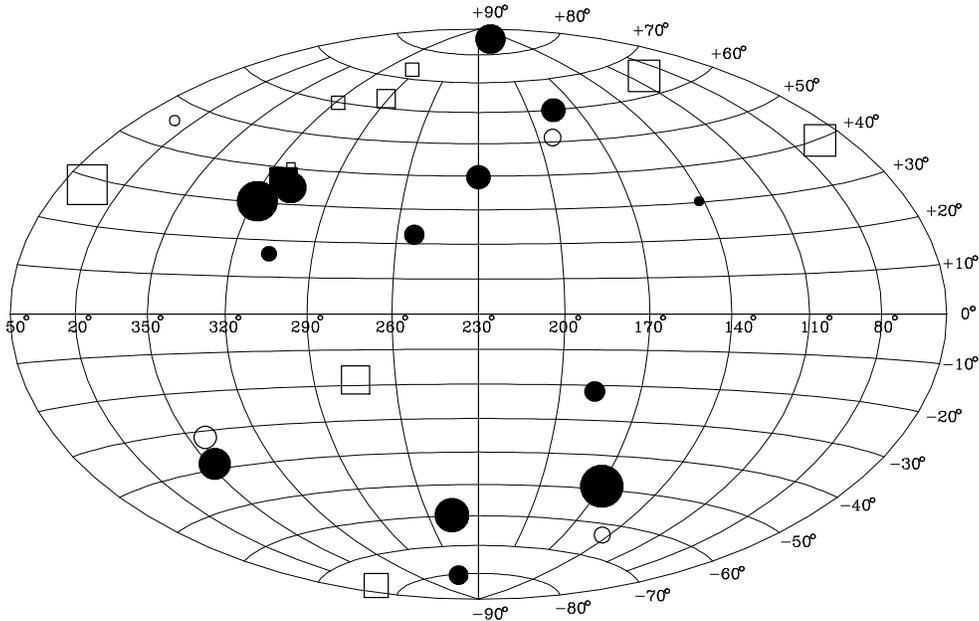}}
\vspace{8.5 cm}
\caption{\capt
The angular positions of the SNe, in galactic coordinates, inside the
void $d<70\hmpc$ (circles) and in the shell around it $70 < d <
120\hmpc$ (squares). Filled and open symbols mark outwards and inwards
peculiar velocities respectively and the size of the symbol is
proportional to the absolute value of the peculiar velocity in the CMB
frame.}
\label{fig:sky}
\end{figure}

% Sky coverage
One concern in the interpretation of our result is that the sparse
sampling and the incomplete sky coverage (especially at low galactic
latitudes) may introduce a bias in the peculiar monopole due to its
covariance with higher multipoles (\eg, Feldman \& Watkins 1994;
1998).  Figure~\ref{fig:sky} is an aitoff plot of the angular
positions of the 27 inner SNe, in the void and in the shell around it,
with their peculiar velocities indicated.

% Bulk Flow
The fact that a similar monopole signal is seen both in the north and
the south Galactic caps suggests that the mixup with a bulk flow is
not likely to be severe.  To quantify the sensitivity of the monopole
to a possible bulk velocity of a similar extent, we redo the 2-zone
model fit with an additional free vector of bulk velocity (in the CMB
frame) within each zone.  We find that $\delta_H$ within $70 \hmpc$
typically becomes lower by only about $1\%$, and the void significance
is reduced.
 
While the monopole can be properly determined out to large distances,
the bulk velocity is much harder to determine at large distances.
This is because the relevant residuals for the monopole are the errors
in $\delta_H$, which are largely independent of distance, whereas the
residuals relevant for the bulk velocity are the errors in the
absolute peculiar velocities, which are generally proportional to
distance.
The best-fit bulk velocity in the void is found to be $(200 \pm 300)
\kms$ (note the large error!), in the general direction of the Local
Group motion relative to the CMB frame and in agreement with a
straightforward extrapolation of the bulk velocity found from galaxy
peculiar velocities in the inner $60\hmpc$ (Dekel \etal 1998), but not
much different from zero.

We have also checked the effect of subtracting out from the data a
bulk flow that is fixed by assumption, based on the extrapolation of
the galaxy bulk flow from the inner $60\hmpc$, namely $200\kms$ in the
direction of the CMB dipole.  Subtracting such a bulk velocity from
the SNe velocities inside the void produces an effect on the void
depth that is similar to the case when the bulk velocity was free in
the fit.  The probability of the outflow monopole being a chance
occurrence is increased from $p=2.2\%$ to about $7\%$ in this case.
 
To investigate further the possible cross-talk with bulk velocity, we
generate synthetic data from a toy model whose ``true" velocity field
is the same as our best-fit 2-zone model, namely a pure Hubble flow
outside $70\hmpc$ and a $6.5\%$ faster Hubble flow inside. We
alternatively add to the ``true" Hubble flow model a constant bulk
velocity.  This toy flow model is ``observed", with the appropriate
errors, at the real positions of the 44 SNe, and our statistical
analysis is applied. The effect of the bulk flow on the solution for
the monopole is found to be weak.  When no bulk flow is present in the
``true" toy model and no bulk velocity is allowed in the model fit, we
find that the void is reproduced properly.  Then, when allowing for a
free bulk velocity in the fit, the correct void is reproduced,
independent of whether a bulk velocity was present in the ``true" toy
model.  However, when an assumed bulk velocity is subtracted out from
the toy data and then a monopole is fitted, the void is reproduced
accurately only when the correct bulk flow is assumed. For example,
when the procedure is applied to a toy model of no ``true" bulk flow
and a $200\kms$ bulk velocity is assumed in the fit, the result
underestimates the depth of the void by $\sim 1\%$.  When there is a
``true" bulk flow in the model which is not allowed in the fit, the
depth is overestimated by $\sim 1\%$.  These tests confirm our
expectation that, although some covariance between monopole and dipole
is inevitable given the specific sampling positions of these 44 SNe,
the effect on the recovered void is small.

% Quadrupole
The possible confusion with a quadrupole moment is worse.  The poor
coverage at low and moderate Galactic latitudes (\eg, at $0 < l < 170$
and $-60 < b < +30$ which includes the Perseus-Pisces region), makes
it practically impossible to distinguish between a peculiar monopole
and a quadrupole.  Future observations which improve the coverage of
regions of moderate Galactic latitudes would help constrain the
quadrupole.
We note that previous investigations of the quadrupole flow (\eg,
Lilje, Yahil \& Jones 1986) have pointed to a pattern that is almost
orthogonal to an outflow at the Galactic poles that would be mistaken
for our monopole.  We also know that the mass distribution in the
Galactic plane is dominated by two big attractors in opposite sides of
the sky, the Great Attractor and the Perseus Pisces supercluster,
which would rather induce an inflow along the Galactic poles.
The current data set is too small for a quantitative analysis of 
possible aliasing by quadrupole and higher multipoles, which is  
left for future analysis with more data.

%==============================================

\section{DISCUSSION}
%5
\label{sec:discussion}
 
% Summary of results
We have analyzed the monopole perturbations in the peculiar velocity
field as probed by 44 SNe Ia with small distance errors.  For an
Einstein-de Sitter universe, we find an outward perturbation in the
Hubble flow of $(6.5 \pm 2.2)\%$ within a sphere of radius $70 \hmpc$,
with a possible indication for inflow in the surrounding shell.  The
void is less significant if $\omm$ is low and $\oml$ is high. Its
significance is increased when one outlier is excluded, or when the
size of the void is constrained \apriori to a limited range.

% Density profiles corresponding to 2-zone and 3-zone models.
In the idealized case of a spherically symmetric density distribution,
an observed monopole in the velocity field is associated with a
specific density profile around the Local Group.  In linear theory,
$\delta \rho/\rho = -\nabla\cdot\vv/\Omega^{0.6}$, so that a constant
velocity monopole $\delta_H$ corresponds, for $\Omega=1$, to a
constant density fluctuation offset of $\delta \rho/\rho = - 3\,
\delta_H$.  In this picture, our results would require a local void of
$\sim -20\%$ density contrast.
 
The sharp drop in $\delta_H$ at $R=70-75\hmpc$ would be produced by a
narrow shell of high mass density at that distance.  This is
approximately the observed location of the shell-like structure
defined in redshift surveys by the Great Wall of Coma and the Southern
Wall. It is tempting to associate the drop in the outflow we see with
these features of the regional galaxy distribution.
Indeed, Dell'antonio, Geller \& Huchra (1996), based on a sample of
peculiar velocities of spiral galaxies in the vicinity of the Great
Wall, estimate the real-space width of the wall to be smaller than
$11\hmpc$ at $90\%$ confidence. Their corresponding limit on the
overdensity of the Great Wall is $\delta \rho/\rho < 2/ \Omega^{0.6}$.
 
An outer region of negative mean $\delta_H$ beyond $75\hmpc$ would
represent back-streaming from voids behind the walls into the walls.
We expect that the density profile averaged over a sphere returns to
its mean value ($\delta M/M=0$) where the velocity monopole converges
to the global Hubble flow.
 
The evidence for such a local void in density profiles derived from
redshift surveys is not conclusive. While we see no evidence for a
local void in the IRAS 1.2 Jy survey as analysed by Koranyi \& Strauss
(1997, Fig.\ 6), Fig.\ 5 of Springel \& White (1997) clearly shows
evidence for a void based on the same data. A similar void is seen in
Saunders \etal (1990, Fig.\ 11) based on the QDOT-IRAS redshift
survey. Preliminary results from the PCSZ-IRAS redshift survey provide
further evidence for the presence of a local void of even larger
extent (S.D.M.\ White 1997, private communication). The apparently
conflicting results could be partly explained by the fact that the
effective survey size, especially in the 1.2 Jy survey, is not much
larger then the volume of the local void. This might introduce some
confusion between the radial selection function and the real density
profile.

Because of the small number of data points in the current sample, we
have not attempted to determine the center point about which the
angle-averaged void is deepest or largest. This center point does not
necessarily coincide with the Local Group.

% CDM models 
Such a void is not in significant conflict with expectations from
standard cosmological theories. The expected \rms perturbations on
these scales, as predicted by popular CDM models, range from $2\%$ to
$4\%$, with standard-CDM ($\omm=1$) being on the high side and Open
CDM ($\omm=0.3$, $\oml=0$) on the low side (see also Shi \& Turner
1998). The local void we find, of $\sim 6 \pm 2\%$, is thus less than
a $2 \sigma$ deviation from the predictions of any of the conventional
models. Although an open model is less likely than the others, the
void in this case is slightly shallower and corresponds only to a
$\sim 1.9\sigma$ deviation from the predictions of an open CDM model.

% Obs. systematics
One should also worry about possible observational systematics that
may be artificially interpreted as a local peculiar outflow.  Sources
of observational error that vary systematically with distance are
prime suspects.  These include the effects of redshift on the
integrated spectral light which passes through a fixed broad band
filter, the ``K-correction''.  Although care has been taken to include
such corrections, the non-stellar features of a SN Ia spectrum inhibit
an exact compensation for this effect.
However, the magnitude of the correction is fairly small for the bulk
of our SNe data, typically less than $0.02$ mag (Hamuy \etal 1993b).
This corresponds to a peculiar velocity of $\sim 50\kms$, which is
less than $1\%$ of deviation from Hubble flow within the local void
compared to the detected signal of $\sim 6\%$ deviation.  Our estimate
of $10\%$ error in the K-correction implies that these effects are
negligible.

Similarly, the range of supernova luminosities observed in the nearby
sample is larger than at great distances because it includes
intrinsically faint objects. Any defects in the ability of MLCS to
correctly account for these systematic shifts could, in principle,
cause problems in making reliable inferences about the void dynamics.
However, this does not seem to be an important effect. The faint tail
in the inner region is populated by only three SNe, of which only one,
the brightest of the three, has a significant positive $\delta H / H$.
 
We have also tested for correlations between the distances or peculiar
velocities and certain quantities that might have systematically
affected them These quantities include the SN detection time with
respect to the time of the light-curve maximum, the absolute
correction to the light-curve template, and the magnitude of the
extinction correction.  We have found no significant correlations.

% Larger void
One implication of our result is that the local value of the Hubble
constant cannot be {\it much} larger than the global value.  Our best
estimate of $\delta_H$ is $(6.5 \pm 2.2)\%$, so if the regional value
(inside $70\hmpc$) were $70\kmsmpc$, the global value would be
$65\kmsmpc$.  But for the global value to reach $50\kmsmpc$ would
require $\delta_H$ of $40\%$, which is clearly excluded by our
analysis.  This assumes that the volume sampled by our set of SNe Ia
reaches out to the mean density and that we are not embedded in a
deeper void of even larger extent.  A possible hint for such a void
may come from the observed over-abundance in faint galaxy counts,
which may, or may not, be explained by luminosity evolution of
galaxies (\eg, Heydon-Dumbelton, Collins, \& McGillivray 1989; Maddox
et al 1990; Lilly 1993; Driver, Windhorst, \& Griffiths 1995; Gronwall
\& Koo 1995).  One controversial interpretation is of a significant
deficiency of galaxies on size scales of $\sim 300\hmpc$ (Huang et al
1997). However, in the Las Campanas Redshift Survey, the galaxy
density profile clearly converges to a universal value beyond
$150\hmpc$ (Lin \etal 1997, Fig. 8).  Furthermore, a comparison of
supernovae distances at low and high redshifts puts an upper limit of
$10\%$ at $95\%$ confidence on the depth of a void of size $\sim
300\hmpc$ (Kim \etal 1997). Our tentative local void detection is also
consistent with upper limits on $\delta_H$ of $\sim 7\%$ obtained from
Abell/ACO clusters (Lauer \& Postman 1992; Lauer \etal 1997).

% Implications
The $\sim 6.5\%$ outflow detected here would partially help reconcile
the large-scale estimates of the Hubble constant, \eg, from SNe
(Tammann \& Sandage 1995; Riess \etal 1996) and estimates that are
limited more to the inner volume, such as based on Tully-Fisher
distances (Kennicutt, Freedman, \& Mould 1995, Mould \etal 1995,
Freedman \etal 1994, Freedman 1997). The latter are typically $\sim
10-20\%$ higher.  The local void marginally detected here would help,
in a humble way, in the resolution of the apparent conflict between
the Hubble constant and the ages of globular clusters (Bolte \& Hogan
1995; Chaboyer \etal 1998).
The implication for the value of $\Omega$ is that estimates based on
data within the local void (see a review by Dekel, Burstein, \& White
1997) would underestimate the universal value by $\sim 20\%$.

We conclude by reiterating that our detection of the local void is
only marginal. Since its statistical significance can be interpreted
in several different ways, we presented a ``supermarket" of
statistical evaluations.
If we were investigating the properties of a void whose scale had been
established by other data, such as the great walls seen in redshift
slices and pencil beams, we would report a $\sim 6.5\%$ outflow which,
in an Einstein de-Sitter universe, has a probability of order $1\%$ to
be obtained by chance from a pure Hubble flow.  The void would be even
more significant if one outlier were removed.
However, with no pre-conception about the extent of the void, the
detection is less decisive, with the probability for a chance
occurrence rising to a few percent.  For the low $\omm$ and high
$\oml$ models this probability is as high as $10\%$, and in every case
there is possible confusion with higher moments of the velocity field,
especially a quadrupole.
In any case, our indication of a modest local void is tentative and
should be confirmed (or refuted) by future observations; our detection
poses a specific model to be tested, that of an outflowing region of
radius $70\hmpc$.

%============================================
\acknowledgments{
We thank George Blumenthal, Bill Press and Xiangdong Shi for
stimulating discussions.  This research was supported in part by the
US-Israel Binational Science Foundation grant 95-00330, by the Israel
Science Foundation grant 950/95, by a NASA Theory grant at UCSC, by
NSF grants AST95-28899 and AST96-17058 at Harvard University, and by
NSF grant PHY94-07194 at the Institute for Theoretical Physics, Santa
Barbara.  }

%=============================================
{}

\end{document}